\documentclass[doublecol]{epl2} 
\usepackage{amsmath,amsthm}
\usepackage{rotating}
\usepackage{blkarray}
\usepackage{multirow}
\usepackage{adjustbox}
\usepackage{graphicx}
\usepackage{mathtools}

\usepackage{amsfonts}
\newtheorem{theorem}{Theorem}

\newtheorem{proposition}[theorem]{Proposition}

\theoremstyle{remark}

\theoremstyle{definition}
\newtheorem{definition}[theorem]{Definition}

\theoremstyle{example}

\theoremstyle{notation}

\newcommand{\bra}[1]{\langle#1|}
\newcommand{\ket}[1]{|#1\rangle}

\begin{document}

\title{Coherent  states with minimum Gini uncertainty for finite quantum systems}            
\author{C. Lei and A. Vourdas}
\institute{Department of Computer Science,\\
University of Bradford, \\
Bradford BD7 1DP, United Kingdom}

\abstract{
Uncertainty relations $\Delta(\rho)\ge \eta_d$ in terms of the Gini index  are studied.
The `Gini uncertainty constant' $\eta_d$ is estimated numerically and compared to an upper bound $\tilde \eta_d\ge \eta_d$.
It is shown that for large $d$ we get $\tilde \eta_d\approx \eta_d$.
States $\ket{g}$ with minimum Gini uncertainty and displacement transformations are used to define 
coherent states $\ket{\alpha, \beta}_g$ (where $\alpha, \beta \in {\mathbb Z}_d$) with minimum Gini uncertainty ($\Delta[\ket{\alpha, \beta}_g\;_g\bra{\alpha, \beta}]\approx \eta_d$).
The $\ket{\alpha, \beta}_g$  resolve the identity, and therefore an arbitrary state can be expanded in terms of them.
This expansion is robust in the presence of noise.
}
\maketitle

\section{1.Introduction}
Uncertainty relations are very important in quantum physics.  For the harmonic oscillator we have the Heisenberg uncertainty relation (and the Robertson-Schr\"odinger uncertainty relation which is related to it).
We also have entropic uncertainty relations (e.g.,\cite{U0}).

For systems with finite-dimensional Hilbert space (e.g.\cite{V1}) that we consider in this paper, we have the entropic uncertainty relations\cite{U1,U2,U3,U4},  
the Donoho-Stark uncertainty relations\cite {U5}, uncertainty relations that use the Gini index\cite{U6}, etc.

The Gini index (e.g.,\cite{G1,G2}) is an alternative to the standard deviation and is popular in Mathematical Economics.
If $\rho$ is a density matrix, uncertainty relations with the Gini index consider the probability distributions ${\cal P}_X(r|\rho)$ and ${\cal P}_P(r|\rho)$ related to dual `position' and `momentum' bases (defined explicitly in Eq.(\ref{62}) below).
Then we calculate the Gini indices 
${\cal G}_X(\rho)$ and ${\cal G}_P(\rho)$ 
for these probability distributions.
It has been shown\cite{U6} that 
\begin{equation}\label{1}
\Delta(\rho)\ge \eta_d
\end{equation}
where $\Delta(\rho)$ is defined in terms of Gini indices in Eq.(\ref{unce}) below, and $\eta_d$ is a constant
(defined in Eq.(\ref{10}) below) which we call `Gini  uncertainty constant' and which depends only on the dimension $d$.
$\eta_d$ is the analogue of $\frac{1}{2}$ in the Heisenberg uncertainty relation for the harmonic oscillator.

In this paper:
\begin{itemize}
\item
Using a large number of pure states, we estimate numerically $\eta_d$ as functions of $d$ (in fig.\ref{f2}).
We also calculate analytically an upper bound 
\begin{eqnarray}
\tilde \eta_d \coloneqq \frac{d-1}{d+1}\cdot \frac{\sqrt{d}}{1+\sqrt{d}}, 
\end{eqnarray}
and we show numerically that for large $d$ we get
$\tilde \eta_d\approx \eta_d$ (in fig.\ref{f3}).
\item
We find numerically states $\ket{g}$ that satisfy the Gini uncertainty inequality as equality ($\Delta(\ket{g}\bra{g})\approx \eta_d$).  
We use these states and displacement transformations, to define 
coherent mixed states $\ket{\alpha, \beta}_g$ (where $\alpha, \beta \in {\mathbb Z}_d$) with minimum Gini uncertainty ($\Delta[\ket{\alpha, \beta}_g\;_g\bra{\alpha, \beta}]\approx \eta_d$).
We show that they are related to each other through displacement transformations, and also that they resolve the identity. 
The resolution of the identity is used for an expansion of an arbitrary state in terms of them (Eq.(\ref{11A})).
It is shown that this expansion is robust in the presence of noise.
\end{itemize}

In section 2 we study a family of states with coherence properties in the sense of proposition \ref{pro1}.
In section 3 we estimate numerically $\eta_d$ and calculate analytically an upper bound $\tilde \eta_d$.
For large $d$, we get $\tilde \eta_d\approx \eta_d$.

In section 4 we  define coherent states $\ket{\alpha, \beta}_g$ with minimum Gini uncertainty.
They resolve the identity, and we can expand an arbitrary state in terms of them.
One of the merits of studying coherent states, is that expansions in terms of them are robust in the presence of noise.
We show that this is the case in the present context.

In section 5 we show that
states that minimise the `Gini uncertainty relation' do not minimise the `entropic uncertainty relation'.
Different uncertainty relations are different constraints for the quantum formalism.

We conclude in section 6 with a discussion of our results.

\section{2.States with coherence properties in finite quantum systems}

We consider a quantum system with variables in ${\mathbb Z}_d$, the ring of integers modulo $d$.
$H_d$ is the $d$-dimensional Hilbert space describing these systems. 
Some of the results in these systems are different for odd or even $d$, and in this paper $d$ is an odd integer.

$|X;r\rangle$ where $r=-\frac{d-1}{2},...,\frac{d-1}{2}$, is an orthonormal basis which we call position basis (the $X$ in this notation is not a variable, but it simply indicates position basis).
Through a Fourier transform we get another orthonormal basis that we call momentum basis:
\begin{eqnarray}
|{P};r\rangle \coloneqq F|{X};r\rangle;\;\;\;\;
F=\frac{1}{\sqrt d}\sum _{r,s}\omega^{rs}\ket{X;r}\bra{X;s};\nonumber\\
\omega=\exp \left (i\frac{2\pi }{d}\right );\;\;\;r,s=-\frac{d-1}{2},...,\frac{d-1}{2}.
\end{eqnarray}

The displacement operators $Z^\alpha, X^\beta$ in the phase space ${\mathbb Z}_d\times {\mathbb Z}_d$, are given by
\begin{eqnarray}\label{2A}
Z^\alpha&\coloneqq& \sum _m\omega^{\alpha m}|X; m\rangle\bra{X;m}\nonumber\\
          &=&\sum _m|P; m+\alpha\rangle \bra{P; m}\nonumber\\
X^\beta &\coloneqq&\sum _m|X; m+\beta\rangle \bra{X; m}\nonumber\\
        &=& \sum _m\omega^{-\beta m}|P; m\rangle\bra{P;m}.
\end{eqnarray}
where $\alpha, \beta \in {\mathbb Z}_d$. 
General displacement operators are the unitary operators
\begin{eqnarray}
&&D(\alpha, \beta)\coloneqq Z^\alpha X^\beta \omega^{-2^{-1}\alpha \beta}\nonumber\\&&[D(\alpha, \beta)]^{\dagger}=D(-\alpha, -\beta);\;\;\;\alpha, \beta \in {\mathbb Z}_d
\end{eqnarray}
The $2^{-1}=\frac{d+1}{2}$ exists in ${\mathbb Z}_d$ because $d$ is an odd integer.
They act on position states as follows;
\begin{eqnarray}\label{62}
D(\alpha, \beta)\ket{X;\kappa}=\omega^{2^{-1}\alpha \beta+\alpha\kappa}\ket{X;\kappa+\beta}
\end{eqnarray}

The $D(\alpha, \beta)\omega ^\gamma$ with $\alpha$, $\beta , \gamma \in {\mathbb Z}_d$, form a representation of the Heisenberg-Weyl group with multiplication rule:
\begin{eqnarray}\label{61}
[D(\alpha_1, \beta_1)\omega ^{\gamma_1}][D(\alpha_2, \beta_2)\omega^{\gamma_2}]\nonumber\\
=D(\alpha, \beta)\omega^\gamma
\end{eqnarray}
where
\begin{eqnarray}
\alpha=\alpha_1+\alpha_2;\;\;\;\beta=\beta_1+\beta_2;\nonumber\\
\gamma=\gamma_1+\gamma_2+2^{-1}(\alpha_1\beta_2-\alpha_2\beta_1)
\end{eqnarray}

If $\ket{f}$ is a state (`fiducial state'), let $\Sigma(\ket{f})$ be the set of $d^2$ density matrices
 \begin{eqnarray}\label{7}
\ket{\alpha, \beta}_f \coloneqq D(\alpha, \beta)\ket{f};\;\;\;\alpha, \beta \in {\mathbb Z}_d,
\end{eqnarray}
The fiducial vector should be a `generic' vector (other than position or momentum states).
They have `coherence properties' described in the following proposition.
\begin{proposition}\label{pro1}
\mbox{}
\begin{itemize}
\item[(1)]
The displacement transformations $D(\kappa, \lambda)\exp(i\mu)$ leave invariant the set $\Sigma(\ket{f})$, in the sense that they transform states in this set into other states in the same set:
\begin{eqnarray}\label{60}
[D(\kappa, \lambda)\omega^\mu]\ket{\alpha, \beta}_f=\omega^\nu\ket{\alpha+\kappa, \beta+\lambda}_f\nonumber\\
\nu=\mu+2^{-1}(\kappa\beta-\lambda\alpha)
\end{eqnarray}
\item[(2)]
The $\ket{\alpha, \beta}_f$ resolve the identity:
\begin{eqnarray}\label{reso}
\frac{1}{d}\sum_{\alpha, \beta}\ket{\alpha, \beta}_f\;_f\bra{\alpha, \beta}={\bf 1}.
\end{eqnarray}

\end{itemize}
\end{proposition}

\begin{proof}
\mbox{}
\begin{itemize}
\item[(1)]
Eq.(\ref{60}) is proved using Eq(\ref{61}).
\item[(2)]
We need to prove that
\begin{eqnarray}
\frac{1}{d}\sum_{\alpha, \beta}\langle X;\kappa \ket{\alpha, \beta}_f\;_f\bra{\alpha, \beta} X;\lambda \rangle=\delta (\kappa, \lambda).
\end{eqnarray}
where $\delta (\kappa, \lambda)$ is the Kronecker delta.
Using Eq.(\ref{62}) this reduces to proving that
\begin{eqnarray}
\frac{1}{d}\sum_{\alpha, \beta}\omega^{\alpha(\kappa-\lambda)}\bra{X;\kappa-\beta} f \rangle\langle f\ket{X;\lambda-\beta}=\delta (\kappa, \lambda).
\end{eqnarray}
But
\begin{eqnarray}
\frac{1}{d}\sum_{\alpha}\omega^{\alpha(\kappa-\lambda)}=\delta (\kappa, \lambda),
\end{eqnarray}
and this reduces to
\begin{eqnarray}
\delta (\kappa, \lambda)\sum_{\beta}|\langle f\ket{X;\kappa-\beta}|^2=\delta (\kappa, \lambda),
\end{eqnarray}
which is true.

\end{itemize}
\end{proof}

Using the resolution of the identity in Eq.(\ref{reso}) we expand an arbitrary state $\ket{s}$ as
\begin{eqnarray}\label{11}
\ket{s}=\sum_{\alpha, \beta}s_{\alpha, \beta}\ket{\alpha, \beta}_f;\;\;\;s_{\alpha, \beta}=\frac{1}{d}\;_f\langle \alpha, \beta \ket{s}.
\end{eqnarray}

The states $\ket{\alpha, \beta}_f$ in Eq.(\ref{7}) have the coherence properties described in proposition \ref{pro1} , and have been studied 
using the language of analytic functions (and in particular Theta functions) in refs.\cite{A1,A2}.
These states are not (in general) minimum uncertainty states.
In this paper we identify among them, which states have minimum quantum uncertainty (defined through Gini uncertainties).

\section{3.Gini uncertainties for finite quantum systems} 

If $\rho$ is a density matrix, the probability distributions related to the position and momentum basis, are
\begin{eqnarray}
&&{\cal P}_X(r|\rho)=\bra{X;r}\rho\ket{X;r};\;\sum _{r=0}^{d-1}{\cal P}_X(r|\rho)=1;\;r\in {\mathbb Z}_d\nonumber\\
&&{\cal P}_P(r|\rho)=\bra{P;r}\rho\ket{P;r};\;\sum _{r=0}^{d-1}{\cal P}_P(r|\rho)=1.
\end{eqnarray}
They are both measurable quantities.

\begin{definition}\label{def56}
The Gini index with respect to the position basis is given by one of the following equivalent to each other relations (e.g., proposition 6.1 in ref\cite{VV}):
\begin{itemize}
\item[(1)]
We use a permutation $\pi_X$ to order the probabilities ${\cal P}_X(r|\rho)$ in ascending order:
\begin{eqnarray}\label{84}
{\cal P}_X(\pi_X(0)|\rho)\le {\cal P}_X(\pi_X(1)|\rho)\le \nonumber\\
...\le {\cal P}_X(\pi_X(d-1)|\rho)]
\end{eqnarray}
Then
\begin{eqnarray}\label{87}
{\cal G}_X(\rho)&=&1-\frac{2}{d+1}[d{\cal P}_X(\pi_X(0)|\rho)\nonumber\\
&+&(d-1){\cal P}_X(\pi_X(1)|\rho)+\nonumber\\
&...&+{\cal P}_X(\pi_X(d-1)|\rho)]
\end{eqnarray}
\item[(2)]
\begin{eqnarray}\label{3d}
{\cal G}_X(\rho)=\frac{1}{2(d+1)}\sum _{r,s}|{\cal P}_X(r|\rho)-{\cal P}_X(s|\rho)|
\end{eqnarray}
\end{itemize}
\end{definition}

Roughly speaking the standard deviation is sum of differences squared, 
while the Gini index is sum of the absolute values of the differences.
The ${\cal G}_X(\rho)$ indicates the Gini uncertainty in ${\cal P}_X(r|\rho)$.
${\cal G}_X(\rho)=\frac{d-1}{d+1}$ indicates a certain outcome (one ${\cal P}_X(r|\rho)$ equal to $1$ and the others equal to zero), while 
${\cal G}_X(\rho)=0$ indicates the most uncertain outcome (${\cal P}_X(r|\rho)=\frac{1}{d}$).
We note that the probabilities ${\cal P}_X(r|\rho)$, and therefore ${\cal G}_X(\rho)$, are measurable quantities.

It has been shown\cite{U6} that
\begin{eqnarray}\label{9}
0\le {\cal G}_X(\rho)\le\frac{d-1}{d+1};\;\;\;{\cal G}_X[\rho(\alpha, \beta)]={\cal G}_X(\rho),
\end{eqnarray}
where 
\begin{eqnarray}
\rho(\alpha, \beta)=D(\alpha, \beta)\rho [D(\alpha, \beta)]^\dagger.
\end{eqnarray}

In a similar way to ${\cal G}_X(\rho)$, we define the Gini index ${\cal G}_P(\rho)$ in terms of the ${\cal P}_P(r|\rho)$ with respect to the momentum basis .
We also define the sum
\begin{eqnarray}\label{23}
{\cal G}_{XP}(\rho)\coloneqq {\cal G}_X(\rho)+{\cal G}_P(\rho).
\end{eqnarray}
Both ${\cal G}_X(\rho)$ and ${\cal G}_P(\rho)$ are measurable quantities using two different ensembles describing the density matrix $\rho$
(because the projectors $\ket{X;r}\bra{X;r}$ and $\ket{P;s}\bra{P;s}$ do not commute).
Therefore ${\cal G}_{XP}(\rho)$ is a measurable quantity.

Ref.\cite{U6} has shown that ${\cal G}_{XP}(\rho)$ cannot take values which are arbitrarily close to $2\frac{d-1}{d+1}$.
Therefore
 \begin{eqnarray}\label{10}
\eta _d \coloneqq 2\frac{d-1}{d+1}-\sup _{\rho }[{\cal G}_{XP}(\rho)]>0.
\end{eqnarray}
We refer to this as the `Gini uncertainty constant', and we get the uncertainty relation
\begin{eqnarray}\label{unce}
&&\Delta(\rho)\coloneqq 2\frac{d-1}{d+1}-{\cal G}_{XP}(\rho)\ge\eta_d.
\end{eqnarray}
Using Eq.(\ref{9}) we show that
\begin{eqnarray}\label{107}
\Delta[\rho(\alpha, \beta)]=\Delta(\rho).
\end{eqnarray}
Also we note that
\begin{eqnarray}
{\cal G}_{XP} \left (\frac{1}{d}{\bf 1}\right )=0;\;\;\;\Delta \left (\frac{1}{d}{\bf 1}\right )=2\frac{d-1}{d+1}.
\end{eqnarray}

If $\rho, \sigma$ are two density matrices, and $p$ a probability then
\begin{eqnarray}\label{35}
{\cal G}_X[p\rho+(1-p)\sigma]\le p{\cal G}_X(\rho)+(1-p){\cal G}_X(\sigma).
\end{eqnarray}
Let $\rho$ be a density matrix and $\lambda_i$, $\ket{e_i}$ its eigenvalues and eigenvectors:
\begin{eqnarray}
\rho=\sum _i \lambda_i \ket{e_i} \bra{e_i};\;\;\;\sum _i\lambda_i=1.
\end{eqnarray}
Then Eq.(\ref{35}) shows  that 
\begin{eqnarray}
&&{\cal G}_X(\rho)\le \sum _i \lambda_i {\cal G}_X(\ket{e_i} \bra{e_i})\nonumber\\
&&{\cal G}_P(\rho)\le \sum _i \lambda_i {\cal G}_P(\ket{e_i} \bra{e_i})
\end{eqnarray}
and therefore
\begin{eqnarray}\label{70}
{\cal G}_{XP}(\rho)\le {\cal G}_{XP}(\ket{e_{\rm max}}\bra{e_{\rm max}})
\end{eqnarray}
where $\ket{e_{\rm max}}$ is the eigenstate with maximum ${\cal G}_{XP}(\ket{e_i}\bra{e_i})$.

\subsection{Numerical estimate of the Gini uncertainty constant $\eta_d$}
We consider a large set $R(d)$ of random pure states. Then
 \begin{eqnarray}\label{67}
\sup _{\rho }[{\cal G}_{XP}(\rho)]\ge \max _{\rho \in R(d)}[{\cal G}_{XP}(\rho)]
\end{eqnarray}
Eq.(\ref{70}) shows that for the supremum we only need to consider pure states.
The random pure states were produced using Qiskit \cite{qiskit}.

Also for the state
\begin{eqnarray}
\rho=\ket{s}\bra{s};\;\;\;\ket{s}=\sqrt{\frac{\sqrt d}{2\sqrt d+2}}[\ket{X;0}+\ket{P;a}]
\end{eqnarray}
where $a,b\in{\mathbb Z}_d$, we find
\begin{eqnarray}\label{100}
{\cal G}_{XP}(\rho )=\frac{d-1}{d+1}\left (1+\frac{1}{1+\sqrt d}\right).
\end{eqnarray}
Therefore
 \begin{eqnarray}\label{67A}
\sup _{\rho }[{\cal G}_{XP}(\rho)]\ge \frac{d-1}{d+1}\left (1+\frac{1}{1+\sqrt{d}}\right),
\end{eqnarray}
and 
 \begin{eqnarray}\label{67B}
\eta_d\le \tilde \eta_d;\;\;\;\tilde \eta_d= \frac{d-1}{d+1}\cdot \frac{\sqrt{d}}{1+\sqrt{d}}.
\end{eqnarray}

\begin{figure}[!htb]
\centering
\includegraphics[width=8cm]{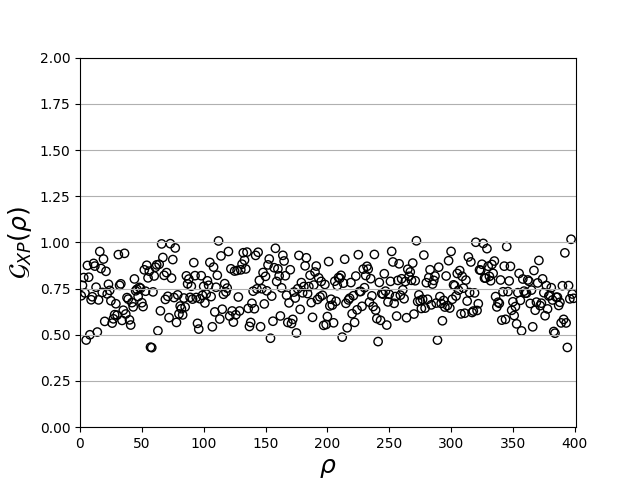}
\caption{${\cal G}_{XP}(\rho)$ for $400$ random pure states ($d=7$) }
\label{f1}
\end{figure}

For $d=7$,  we plot in fig.\ref{f1} the values of ${\cal G}_{XP}(\rho)$ for $400$ random pure states.
The plot shows that  $400$ random states are enough for a very good estimate of the $\sup _{\rho }[{\cal G}_{XP}(\rho)]$.
We note that since we use a random states, every time we repeat the calculation we get a very similar (but not exactly the same) value for $\max _{\rho \in R(d)}[{\cal G}_{XP}(\rho)]$.
We also note that the result obeys the inequality in Eq.(\ref{67A}).

\begin{figure}[!htb]
\centering
\includegraphics[width=8cm]{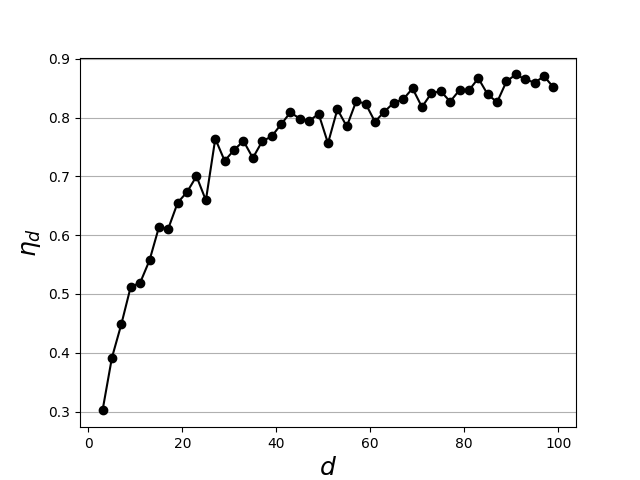}
\caption{$\eta_d$ as a function of (odd) $d$.}
\label{f2}
\end{figure}

Using $400$ random pure states for each value of the odd dimension $d$, we calculated $\max _{\rho \in R(d)}[{\cal G}_{XP}(\rho)]$ (which is a lower bound for $\sup _{\rho }[{\cal G}_{XP}(\rho)]$), and in turn a numerical estimate of $\eta_d$, for odd values of $d$.
The results are shown in fig.\ref{f2} (the `wiggles' are an artefact of the numerical work).

\begin{figure}[!htb]
\centering
\includegraphics[width=8cm]{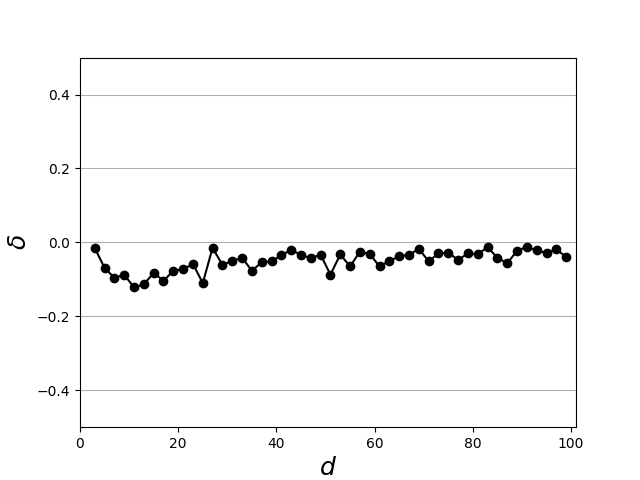}
\caption{$\delta=\eta_d-\tilde \eta_d$ as a function of (odd) $d$. }
\label{f3}
\end{figure}

For large $d$ (e.g. $d>41$) fig.\ref{f3} shows that 
\begin{eqnarray}
\delta \coloneqq \eta_d-\tilde \eta_d\approx 0.
\end{eqnarray}

\section{4.Coherent states with minimum Gini uncertainty}
Let $\ket {g}$ be a state that satisfies (approximately) the uncertainty inequality in Eq.(\ref{unce}) as an equality 
($\Delta(\ket{g}\bra{g})\approx \eta_d$).
Examples of such states for $d=3,5,7$ are given in table \ref{t0}.

\begin{table}[!htb]
 \centering
 \begin{math}
\begin{blockarray}{cc}
   &  &  \\\cline{1-2}
\begin{block}{|c|c|}
$$\vert g\rangle, d =3$$ & \begin{pmatrix}
-0.2395 +0.1773i\\
-0.3749-0.7941i\\
0.3735+0.0232i
\end{pmatrix}\\\cline{1-2}
 $$\vert g\rangle, d=5$$ & \begin{pmatrix}
-0.1665+0.2964i\\
0.5752-0.5821i\\
-0.3445 +0.2167i\\
0.1598+0.0086i\\
-0.1092-0.1072i
\end{pmatrix} \\ \cline{1-2}
 $$\vert g\rangle, d=7$$ & \begin{pmatrix}
0.3479-0.0613i\\
-0.1256-0.0417i\\
-0.0054+0.8010i\\
0.1875+0.1370i\\
-0.1618-0.1764i\\
-0.3214+0.0283i\\
-0.0125+0.0228i
\end{pmatrix} \\ \cline{1-2}
\end{block}
\end{blockarray}
 \end{math}
 \caption{The minimum Gini uncertainty state $\ket{g}$ for $d=3,5,7$.}
 \label{t0}
 \end{table}

According to Eq.(\ref{107}), the $d^2$ states 
 \begin{eqnarray}\label{27}
\ket{\alpha, \beta}_g=D(\alpha, \beta)\ket{g}
\end{eqnarray}
also have minimum Gini uncertainty ($\Delta[\ket{\alpha, \beta}_g\;_g\bra{\alpha, \beta}]\approx \eta_d$).
In addition to that   these states
also obey proposition \ref{pro1}, i.e., they are related to each other through displacement transformations, and they resolve the identity.
Consequently we can expand an arbitrary state  in terms of these states (as in Eq.(\ref{11})):
\begin{eqnarray}\label{11A}
\ket{s}=\sum_{\alpha, \beta}s_{\alpha, \beta}\ket{\alpha, \beta}_g;\;\;\;s_{\alpha, \beta}=\frac{1}{d}\;_g\langle \alpha, \beta \ket{s}.
\end{eqnarray}
As an example we take $d=3$ and 
  \begin{eqnarray}\label{40}
\ket{s}=\begin{pmatrix}
0.5040-0.1526i\\
0.3283+0.1757i\\
0.8324+0.0231i
\end{pmatrix}
\end{eqnarray}  
The $9$ component vectors are given in table \ref{t1}.

There is redundancy in coherent states (in the present context we have $d^2$ coherent states in a $d$-dimensional Hilbert space).
This redundancy makes the expansion in terms of coherent states, robust in the presence of noise.
We show this for the above example.

There are many ways of adding noise to the exact result.
Here we multiplied each of the components $s_{\alpha, \beta}$ in table \ref{t1}, with $1+\lambda_{\alpha, \beta}$ where 
$\lambda_{\alpha, \beta}$ is a random number uniformly distributed in the interval  $(-\epsilon,\epsilon)$:
 \begin{eqnarray}\label{38A}
\ket{\tilde s}&=&\sum_{\alpha, \beta}(1+\lambda_{\alpha, \beta})s_{\alpha, \beta}\ket{\alpha, \beta}_g\end{eqnarray}  
We then calculated the `error vector'
 \begin{eqnarray}
\ket{e}\coloneqq \ket{\tilde s}-\ket{s}=\sum_{\alpha, \beta}\lambda_{\alpha, \beta}s_{\alpha, \beta}\ket{\alpha, \beta}_g
\end{eqnarray}  
and its magnitude
 \begin{eqnarray}
||\ket{e}||=\sqrt{\sum_i|e_i|^2}
\end{eqnarray}  
We repeated that numerical experiment $10$ times and found an average value of $||\ket{e}||_{\rm av}=0.01116$ for $\epsilon=0.3$, and $||\ket{e}||_{\rm av}=0.02095$ for $\epsilon=0.5$.
It is seen that the expansion is robust in the presence of noise.

  \begin{figure}[!htb]
\centering
\includegraphics[width=8cm]{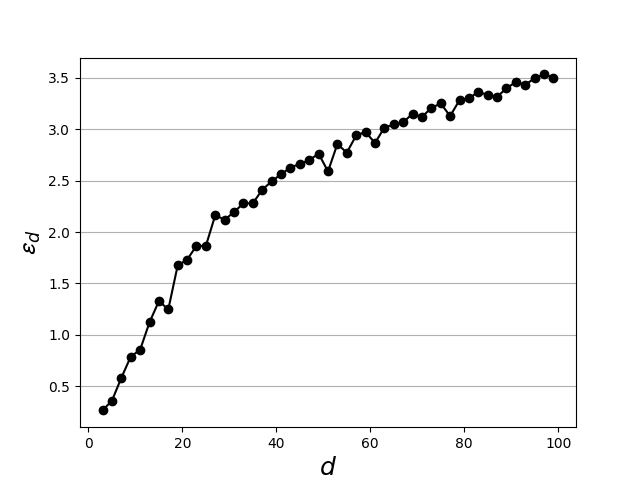}
\caption{${\cal E}_d$ in Eq.(\ref{47}) as a function of (odd) $d$}
\label{f4}
\end{figure} 
 
  \section{5.Comparison with the entropic uncertainty relation}

We consider the Shannon entropies
 \begin{eqnarray}
 E_X(\rho)&=&-\sum _{r=0}^{d-1}{\cal P}_X(r|\rho)\ln {\cal P}_X(r|\rho);\nonumber\\
 E_P(\rho)&=&-\sum _{r=0}^{d-1}{\cal P}_P(r|\rho)\ln {\cal P}_P(r|\rho)
 \end{eqnarray}
which take values in the interval $(0, \ln d)$. Large entropies correspond to large uncertainties.
The entropic uncertainty relation \cite{U1,U2,U3,U4} proves that
 \begin{eqnarray}
{\mathfrak E}(\rho) \coloneqq E_X(\rho)+E_P(\rho)-\ln d\ge 0.
 \end{eqnarray}
 We use logarithms with base $e$.
 
Using the density matrices $\ket{g}$ defined earlier, we calculated the 
 \begin{eqnarray}\label{47}
 {\cal E}_d \coloneqq {\mathfrak E}[\ket{0,0}_g\;_g\bra{0,0}]
  \end{eqnarray}
 In fig.\ref{f4} we plot the $ {\cal E}_d$ as a function of $d$.
It is seen that the states  $\ket{0,0}_g$ 
which are states with minimum Gini uncertainty ($\Delta[\ket{0,0}_g\;_g\bra{0,0}]\approx \eta_d$),
are {\bf not} states with minimum entropic uncertainties.

Different uncertainty relations are different constraints for the quantum formalism.

  \section{6.Discussion}
We studied uncertainty relations for systems with finite-dimensional Hilbert space, using the Gini index (Eq.(\ref{unce})).
They involve the quantity ${\cal G}_{XP}$ in Eq.(\ref{23}), which as we explained earlier is a measurable quantity.
We estimated numerically $\eta_d$, as a function of $d$.
We also gave the analytical upper bound $\tilde \eta_d$ and shown that for large $d$, we get $\tilde \eta_d\approx \eta_d$.

We found numerically states $\ket{g}$ that satisfy the Gini uncertainty inequality, as an equality ($\Delta(\ket{g}\bra{g})\approx \eta_d$).
We used these states to define the
coherent states $\ket{\alpha, \beta}_g$ (Eq.(\ref{27})) that have minimum Gini uncertainty.
They resolve the identity, and  an expansion of an arbitrary state in terms of them is given in Eq.(\ref{11A})).
It has been shown with numerical examples, that this expansion is robust in the presence of noise.

 \begin{table}[!htb]
 \centering
 \begin{math}
\begin{blockarray}{cc}
   &  &  \\\cline{1-2}
\begin{block}{|c|c|}
$$s_{-1,-1}\vert -1,-1\rangle_g$$ & \begin{pmatrix}
0.2527-0.0295i\\ 
0.0844+0.0181i\\
0.1074-0.0148i
\end{pmatrix}\\\cline{1-2}
$$s_{-1,0}\vert -1,0\rangle_g$$& \begin{pmatrix}
0.0893+0.0016i\\
0.2093+0.0081i\\
0.0664+0.0255i
\end{pmatrix} \\ \cline{1-2}
$$s_{-1,1}\vert -1,1\rangle_g$$ & \begin{pmatrix}
0.0923 +0.0310i\\
0.1222 -0.0029i\\
0.2869-0.0008i
\end{pmatrix} \\ \cline{1-2}
$$s_{0,-1}\vert 0,-1\rangle_g$$ & \begin{pmatrix}
0.0672-0.0952i\\
-0.0361-0.0161i\\
0.0217+0.0447i
\end{pmatrix} \\ \cline{1-2}
$$s_{0,0}\vert 0,0\rangle_g$$ & \begin{pmatrix}
-0.0338 -0.0055i\\  
0.0270 +0.0757i\\
0.0234-0.0140i
\end{pmatrix} \\ \cline{1-2}
$$s_{0,1}\vert 0,1\rangle_g$$& \begin{pmatrix}
-0.0108-0.0733i\\
-0.0488+0.0791i\\
0.2180+0.0109i
\end{pmatrix} \\ \cline{1-2}
$$s_{1,-1}\vert 1,-1\rangle_g$$& \begin{pmatrix}
0.0688+0.0071i\\
-0.0191+0.0136i\\
-0.0126-0.0266i
\end{pmatrix} \\ \cline{1-2}
$$s_{1,0}\vert 1,0\rangle_g$$& \begin{pmatrix}
0.0151-0.0175i\\
0.0169+0.0517i\\
-0.0159-0.0094i
\end{pmatrix} \\ \cline{1-2}
$$s_{1,1}\vert 1,1\rangle_g$$& \begin{pmatrix}
-0.0367+0.0287i\\
-0.0274-0.0517i\\
0.1370+0.0078i
\end{pmatrix} \\  \cline{1-2}
\end{block}
\end{blockarray}
 \end{math}
 \caption{The vectors in the expansion in Eq.(\ref{11A}) for the state in Eq.(\ref{40})}
 \label{t1}
 \end{table}

\end{document}